# A Science Model Driven Retrieval Prototype[1]

**Philipp Mayr**
GESIS – Leibniz-Institute for the Social Sciences
Lennéstr. 30, 53113 Bonn, Germany
philipp.mayr@gesis.org

**Philipp Schaer**
GESIS – Leibniz-Institute for the Social Sciences
Lennéstr. 30, 53113 Bonn, Germany
philipp.schaer@gesis.org

**Peter Mutschke**
GESIS – Leibniz-Institute for the Social Sciences
Lennéstr. 30, 53113 Bonn, Germany
peter.mutschke@gesis.org

**ABSTRACT**
This paper is about a better understanding on the structure and dynamics of science and the usage of these insights for compensating the typical problems that arises in metadata-driven Digital Libraries. Three science model driven retrieval services are presented: co-word analysis based query expansion, re-ranking via Bradfordizing and author centrality. The services are evaluated with relevance assessments from which two important implications emerge: (1) precision values of the retrieval service are the same or better than the *tf-idf* retrieval baseline and (2) each service retrieved a disjoint set of documents. The different services each favor quite other – but still relevant – documents than pure term-frequency based rankings. The proposed models and derived retrieval services therefore open up new viewpoints on the scientific knowledge space and provide an alternative framework to structure scholarly information systems.

## 1   INTRODUCTION

In typical metadata-driven Digital Libraries three major difficulties arise: (1) the vagueness between search and indexing terms, (2) the information overload by the amount of result records obtained by information retrieval (IR) systems, and (3) the problem that pure term frequency based rankings, such as term frequency – inverse document frequency (*tf-idf*), provide results that often do not meet user needs (Mayr et al. 2008).

We will present an overall approach to use computational science models as enhanced search stratagems (Bates 1990) within a scholarly IR environment. These computational models can be implemented within scholarly information portals. We assume that a user's search should improve by using these science model driven search tactics when interacting with a scientific information system.

This paper will at first introduce three scientific models: (1) co-word analysis and the derived concept of search term recommendation, (2) coreness of journals and (3) centrality of authors. The basic assumptions and concepts are presented (see in detail Mayr et al. 2011). The section on implementation deals with the set up prototype system that operationalize the three models. The conducted evaluation and study with 73 students is described in the following section. The paper closes with a discussion of the observed results and the conclusion on the presented models.

## 2   MODELS FOR INFORMATION RETRIEVAL ENHANCEMENT

The standard model of IR is the *tf-idf* model which proposes a text-based relevance ranking (Manning et al. 2008). As *tf-idf* is text-based it assigns a weight to term *t* in document *d* which is influenced by different occurrences of *t* and *d*. Variations of the basis term weighing process have been proposed, like normalization of document length or by scaling the *tf* values but the basic assumption stays the same.

### 2.1   Query Expansion via Search Term Recommendation

Search Term Recommenders (STR) are an approach to compensate the long known language problem in Information Retrieval (Blair 2003; Petras 2006): Searching an information system a user has to come up with the "correct" query terms so that they best match the document language to get an appropriate result.

STR are based on statistical co-word analysis and build associations between free terms (i.e. from title or abstract) and controlled terms (i.e. from a thesaurus) which are used during a professional indexation of the documents (see Fig 1.). The co-word analysis implies a semantic association between the free and the controlled terms. The more often terms co-occur in the text the more likely it is that they share a semantic relation. These relations can be calculated and operationalized by different algorithms like LSA, PLSA, SVM and many others. In our setup we use STR for automatic query expansion where the original query of the user is enhanced with semantical "near" terms from a controlled vocabulary.

---
[1] Paper for the Cologne Conference on Interoperability and Semantics in Knowledge Organization.

**Figure 1: Mapping between user terms and controlled terms in the IRM prototype. Example search term "luhmann" and highly associated controlled terms (controlled context on the right) from a STR build on the Thesaurus Sozialwissenschaften.**

## 2.2 Bradfordizing

Bradfordizing is an alternative mechanism to re-rank result lists according to core journals to bypass the problem of very large and unstructured result sets (see Fig. 2). The approach of Bradfordizing is to use characteristic concentration effects (Bradford's law of scattering) that appear typically in journal literature. Bradfordizing defines different zones of documents which are based on the frequency counts in a given document set. Documents in core journals – journals which publish frequently on a topic – are ranked higher than documents which were published in journals from the following zones. In Information Retrieval a positive effect on the search result can be assumed in favor of documents from core journals (White 1981; Mayr 2009).

**Figure 2: Using core journals for reranking in the IRM prototype. Search term "luhmann" and core journals (journal context on the right).**

## 2.3 Author Centrality

Author centrality is another way of re-ranking result sets. Here the concept of centrality in a network of authors is an additional approach for the problem of large and unstructured result sets. The intention behind this ranking model is to make use of knowledge about the interaction and cooperation behavior in special fields of research (see Fig. 3). The (social) status and strategic position of a person in a scientific community is used too. The model is based on a network analytical view on a field of research and differs greatly from conventional text-oriented ranking methods like *tf-idf*.

A concrete criterion of relevance in this model is the centrality of authors from retrieved publications in a co-authorship network. The model calculates a co-authorship network based on the result set to a specific query. Centrality of each single author in this network is calculated by applying the betweenness measure and the documents in the result set are ranked according to the betweenness of their authors so that publications with very central authors are ranked higher in the result list (Mutschke 2004).

Figure 3: Using author centrality for reranking in the IRM prototype. Search term "luhmann" and central authors (author context on the right).

## 3  IMPLEMENTATION

All proposed models are implemented in a live information system using (1) the Solr search engine, (2) Grails Web framework and (3) Recommind Mindserver to demonstrate the general feasibility of the approaches. Solr is an open source search platform from the Apache Lucene project[2] which uses a *tf-idf* based ranking mechanism[3]. The Mindserver is a commercial text categorization tools which was used to generate the STR. Both Bradfordizing and author centrality as re-rank mechanism are implemented as plugins to the open source web framework Grails. Grails is the glue to combine the different modules and to offer an interactive web-based prototype[4].

These retrieval services can be applied in different query phases. In a typical search scenario a user first formulates his/her query, which can then be enriched by a STR that adds controlled descriptors from the corresponding document language to the query. With this new query a search in a database can be triggered. The search returns a result set which can be re-ranked using either Bradfordizing or author centrality. Since search is an iterative procedure this workflow can be repeated many times till the expected result set is retrieved.

## 4  EVALUATION

By measuring the contribution of our services to retrieval performance we expect deeper insights in the structure and the functions of the science system. The evaluation plays the role of a "litmus test" for the adequacy of the science models proposed.

---

[2] http://lucene.apache.org/Solr/

[3] http://lucene.apache.org/java/2_4_0/scoring.html

[4] http://www.gesis.org/beta/prototypen/irm/

## 4.1 Methods

The standard approach to evaluate Information Retrieval systems is to do relevance assessments. In respect to a defined information need documents are marked as relevant or not relevant. Large standard test collections (like TREC, CLEF etc.) contain pre-assessed documents from domain experts where all containing documents are judged relevant or not relevant. Since modern collections usually are too large to be assessed in total only subsets of the collection are assessed. *Pooling* is used to disguise the origin of the document (Voorhees and Harman 2005). In this standard approach subsets of the collections are formed by pooling the top *n* documents returned by the different IR systems to be evaluated. In the next step the assessors have to judge the documents in the subsets without knowing the originating IR systems.

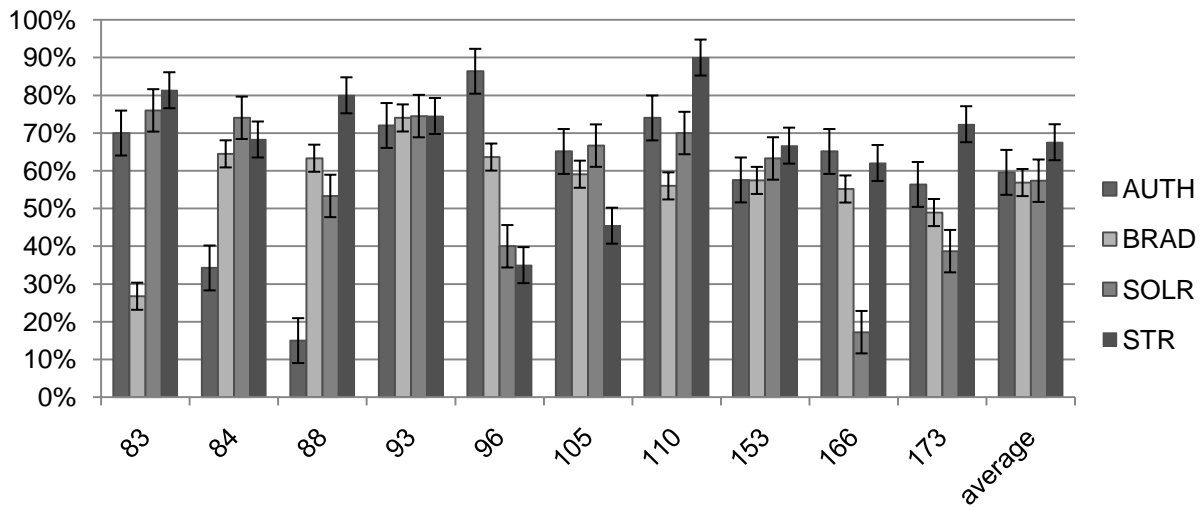

**Figure 4: Precision for each topic and service (Relevance assessments per topic / total amount of single assessments), including standard error**

## 4.2 Assessed Data Set

We conducted a user assessment with 73 students who used the SOLIS database with 369,397 single documents (including title, abstract, controlled keyword etc.). After a briefing each student had to choose one out of 10 different predefined topics. Topic title and the description were presented to form the information need. The pool was formed out of the top n=10 ranked documents from each service and the initial *tf-idf* ranked result set respectively. Duplicates were removed, so that the size of the sample pools was between 34 and 39 documents each. The assessors could choose to judge relevant or not relevant (binary decision) – in case they didn't assess a document this document is ignored in later calculations.

The assessors did 43.78 assessments in average which sums up to 3,196 single relevance judgments in total. Only 5 participants didn't fill out the assessment form completely, but 13 did more than one. Since every assessor could freely choose from the topics the assessments are not distributed evenly. Topic 83 was picked 16 times – topic 96 twice.

# 5 RESULTS

## 5.1 Precision

The precision *P* of each service was calculated by

$$P = \frac{|r|}{|r+nr|}$$

for each topic, where |r| is the number of all relevant assessed documents and |r+nr| is the number of all assessed documents (relevant and not relevant). All precision values and numbers of relevance assessments can be seen in table 1. A graph of all precision values including standard error can be seen in figure 4.

The average precision of the STR was highest (68%) compared to the baseline from the SOLR system (57%). The two alternative ranking methods Bradfordizing (BRAD) and author centrality (AUTH) scored 57% and 60% respectively.

## 5.2 Inter-rater agreement

The assessors in this experiment were not professionals and/or domain experts but mainly library and information science students. However Al-Maskari et al. (2008) showed that this evaluation setup can produce relevant data. They compared official TREC with non-TREC assessors. The agreement changed due to the different topics and the actual ranking position of the assessed document and was between 75 and 82%. The agreement rates of these studies can be compared to our assessments where the overall agreement between all topics and all participants was 82%. 124 of 363 cases were perfect matches where all assessors agreed 100% (all relevant and non relevant judgments matched).

To further rate the reliability and consistency of agreement between the different assessments a statistical measure is needed. Fleiss's Kappa is a measure of inter-grader reliability for nominal or binary ratings (Fleiss 1971). It is an extension of Cohen's Kappa for two-grader ratings. It can be interpreted as expressing the extent to which the observed amount of agreement among raters exceeds what would be expected if all raters made their ratings completely randomly.

Kappa scores can range from <0 (less than chance) over 0.0 (chance) to 1.0 (full agreement). Different interpretations of Fleiss's Kappa were proposed (Landis and Koch 1977 or Osman et al. 2010). All Kappa scores in our experiment range between 0.20-0.40 (topics 84, 110 and 153) and 0.40-0.52 (rest of topics) respectively. Average score was 0.4 and median score was 0.43 which are fair up to moderate levels of agreement or mainly acceptable in the more conservative interpretations.

| id | non relevant | | | | relevant | | | | precision (in %) | | | |
|---|---|---|---|---|---|---|---|---|---|---|---|---|
| | AUTH | BRAD | SOLR | STR | AUTH | BRAD | SOLR | STR | AUTH | BRAD | SOLR | STR |
| 83 | 42 | 104 | 36 | 25 | 98 | 38 | 114 | 109 | 70,00 | 26,76 | 76,00 | 81,34 |
| 84 | 71 | 38 | 27 | 26 | 37 | 69 | 77 | 56 | 34,26 | 64,49 | 74,04 | 68,29 |
| 88 | 51 | 22 | 28 | 12 | 9 | 38 | 32 | 48 | 15,00 | 63,33 | 53,33 | 80,00 |
| 93 | 28 | 26 | 25 | 26 | 72 | 74 | 73 | 76 | 72,00 | 74,00 | 74,49 | 74,51 |
| 96 | 3 | 8 | 12 | 13 | 19 | 14 | 8 | 7 | 86,36 | 63,64 | 40,00 | 35,00 |
| 105 | 15 | 18 | 15 | 24 | 28 | 26 | 30 | 20 | 65,12 | 59,09 | 66,67 | 45,45 |
| 110 | 13 | 22 | 15 | 5 | 37 | 28 | 35 | 45 | 74,00 | 56,00 | 70,00 | 90,00 |
| 153 | 42 | 40 | 36 | 32 | 57 | 54 | 62 | 64 | 57,58 | 57,45 | 63,27 | 66,67 |
| 166 | 30 | 39 | 72 | 33 | 56 | 48 | 15 | 54 | 65,12 | 55,17 | 17,24 | 62,07 |
| 173 | 41 | 48 | 57 | 26 | 53 | 46 | 36 | 68 | 56,38 | 48,94 | 38,71 | 72,34 |
| avg. | | | | | | | | | *59,58* | *56,89* | *57,37* | *67,57* |

**Table 1. Relevance judgments for each topic and service with calculated precision**

## 5.3 Overlap of top document result sets

A comparison of the intersection of the relevant top 10 document result sets between each pair of retrieval service shows that the result sets are nearly disjoint. 400 documents (4 services * 10 per service * 10 topics) only had 36 intersections in total. Thus, there is no or very little overlap between the sets of relevant top ranked documents. The largest, but still very low overlap is among the standard *tf-idf* ranking from SOLR and STR which have 14 common documents.

## 6   DISCUSSION

The evaluation of the STR enhanced system showed that term suggestions can provide an overview over different areas of discussion by adding new concepts. Additionally the STR can support different domains and perspectives (see Petras 2006). This is known as a "query drift" in automatic query expansion but in the application of the STR the service retrieves more relevant documents. While the result set of a STR enhanced query grows and broadens (because of the OR-ing of the added terms) the first n=10 hits are more precise and narrowed. This contradiction can be explained with the high descriptive power of the controlled terms that are added to the query and is an indicator for the high quality of the semantic mapping between the language of scientific discourse (free text in title and abstract) and the language of documentation (controlled thesauri terms).

Discussing the results of the two proposed re-ranking methods Bradfordizing and author centrality brings up two central insights: (1) users get new result cutouts with other relevant documents which are not listed in the first section (first n=10 documents) of the original list and (2) Bradfordizing and author centrality can be a helpful information service to positively influence the search process, especially for searchers who are new on a research topic and don't know the main publication sources or the central actors in a research field.

The Bradford approach always runs the risk and critic of disregarding important developments outside the core. It is often criticized to favor majority views and mainstream journals and ignores minority standpoints. This is a serious argument but it has to be seen as a problem of the data set producers because Bradfordizing only works with existing document sets, which are compiled (and prefiltered) by database producers.

The basic assumption of author centrality based ranking is that central authors are strongly associated with the mainstream topics of a research field and that the phenomenon of authors of high betweenness are supposed to be of high community driving relevance for the science system can be utilized. This perception of the strategic role of highly central actors in science might explain the high precision of rankings done by author betweenness: Authors of high betweenness address the key topics of a field.

## 7   CONCLUSION

Looking at the precision values and the overlap of the result sets two important insights can be noted: (1) precision values of the retrieval services are the same or better than the *tf-idf* retrieval baseline and (2) each service retrieved a disjoint set of documents. The different services each favor quite other – but still relevant – documents than pure term-frequency based rankings. The proposed models and derived r services open up new viewpoints on the scientific knowledge space and also provide an alternative framework to structure the science system.

In a next step we plan to evaluate the proposed retrieval services with strictly scientific topics and scientist who are assessing documents in their specific research field.

## ACKNOWLEDGMENTS

Special thanks go to the students at Humboldt University (guided by Vivien Petras) and University of Applied Science in Darmstadt who took part in our assessment. We thank Hasan Bas who implemented the assessment tool.

This work was funded by DFG, grant no. INST 658/6-1.